\documentclass[final,3p,times,twocolumn]{elsarticle}
 \biboptions{comma,sort&compress}
\usepackage{here}
 \usepackage{graphicx}
  \usepackage{epsfig}

\def\beq{\begin{equation}}
\def\eeq{\end{equation}}
\def\bea{\begin{eqnarray}}
\def\eea{\end{eqnarray}}

\journal{Nuc. Phys. (Proc. Suppl.)}

\usepackage{relsize}
\usepackage{amsmath}
\usepackage{wrapfig}  
\usepackage{amssymb} 
\def\babar{\mbox{\slshape B\kern-0.1em{\smaller A}\kern-0.1em B\kern-0.1em{\smaller A\kern-0.2em R }}} 

\begin{document}

\begin{frontmatter}



\title{Charmonium-like states at \babar}

 \author[label1]{Elisa Fioravanti\corref{cor1}}
  \address[label1]{INFN Ferrara, Via Saragat 1 44122 Ferrara Italy
\\}
\cortext[cor1]{Speaker}
\ead{fioravanti@fe.infn.it}



\begin{abstract}
\noindent
We present new results on charmonium-like states from the \babar experiment located at the
PEP-II asymmetric energy $e^+e^-$ collider at the SLAC National Accelerator Laboratory.
\end{abstract}


\end{frontmatter}



\section{Charmonium Spectroscopy}

The charmonium spectrum consists of eight narrow states below the open charm
threshold (3.73 GeV/c$^2$) and several tens of states above the threshold. Below the threshold almost
all states are well established. On the other hand, very little is known above the threshold, there are several new "Charmonium-like" states that are very difficult to accommodate in the charmonium
spectrum. \\
\indent The B-factories are an ideal place to study charmonium since charmonium states are
produced in four different processes: 
\begin{itemize}
\item B decays, charmonium states of any quantum numbers can be produced 
\item Two photon production: in this process two virtual photons are emitted by the colliding $e^+e^-$ pair ($e^+e^-\rightarrow e^+e^-\gamma^*\gamma^*\rightarrow e^+e^-(c\bar{c}$)), charmonium states with $J^{PC}=0^{\pm+}, 2^{\pm+}, 4^{\pm+}, ... , 3^{++}, 5^{++}...$ can be produced.
\item Initial State Radiation (ISR): where a photon is emitted by the incoming electron or positron ($e^+e^-\rightarrow\gamma c\bar{c}$), only states with $J^{PC}=1^{--}$ are formed
\item Double charmonium production: in this process a $J/\psi$ or a $\psi(2S)$ is produced together with another charmonium state. 
\end{itemize}

\section {Study of the process $\gamma\gamma\rightarrow J/\psi\omega$}
\begin{figure}[htb]
  \centering
  \includegraphics[width=80mm]{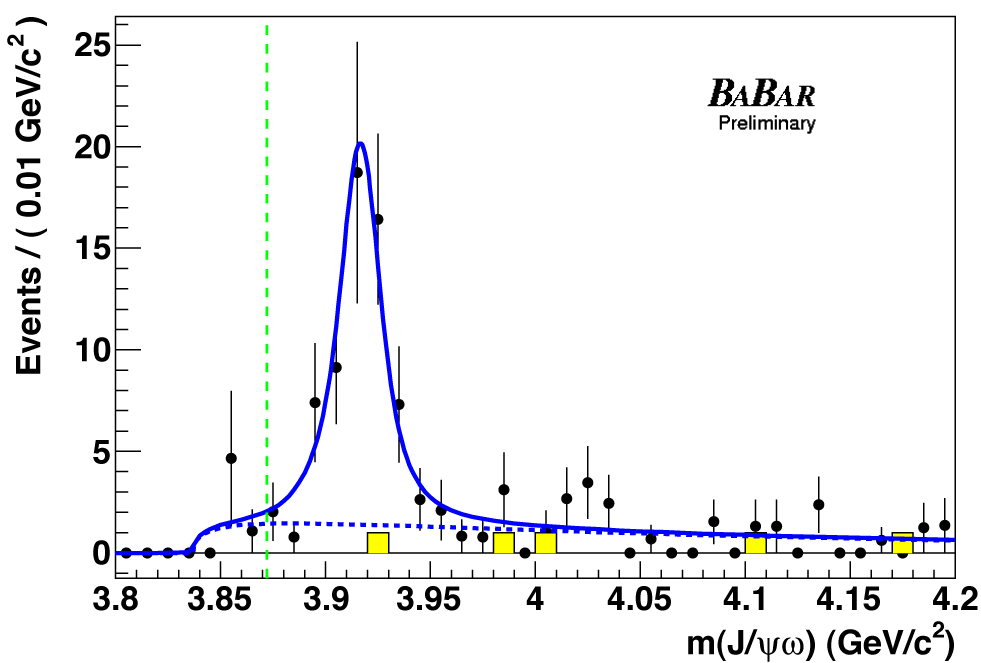}
  \caption{\scriptsize The efficiency-corrected invariant mass distribution for the $J/\psi\omega$ final state. The vertical dashed line is at the X(3872) mass.}
  \label{Fig:Xres1}
\end{figure}

The charmonium-like state Y(3940) was first seen by Belle \cite{Xbelleref2} and then confirmed by \babar \cite{babarref} in the same B meson decay mode, but with lower mass and smaller width compared to the Belle results.\\
\indent In a re-analysis \cite{reana} of the \babar data which used the complete $\Upsilon(4S)$ data sample, the precision of the Y(3940) measurements was improved and evidence for the decay X(3872)$\rightarrow J/\psi\omega$ was reported. This confirmed an earlier unpublished Belle claim for the existence of this decay mode \cite{unpub}. The latter was based on the behaviour of the invariant $\pi^+\pi^-\pi^0$ mass distribution near the X(3872), whereas the \babar result is obtained directly from a fit the the $J/\psi\omega$ mass distribution.\\
\indent A subsequent paper from Belle \cite{belleX3915} reports the observation in $\gamma\gamma\rightarrow J/\psi\omega$ of a state, the X(3915), with mass and width values similar to those obtained for the Y(3940) in the \babar analysis \cite{babarref}.\\
\indent The \babar analysis of the process $\gamma\gamma\rightarrow J/\psi\omega$ has been performed in order to search for the X(3915) and the X(3872) resonances via the decay to $J/\psi\omega$, using a data sample of 519 fb$^{-1}$. Figure \ref{Fig:Xres1} presents the reconstructed $J/\psi\omega$ invariant mass distribution after all the selection criteria have been applied. We perform an extended maximum likelihood fit to the efficiency-corrected spectrum.  A large peak at near 3915 MeV/c$^2$ is observed with a significance of 7.6 $\sigma$. The measured resonance parameters are $m[X(3915)]=(3919.4\pm2.2\pm1.6)$ MeV/c$^2$, $\Gamma[X(3915)]=(13\pm6\pm3)$ MeV. The measured value of the two-photon width times the branching fraction, $\Gamma_{\gamma\gamma}[X(3915)]$ x $\cal{B}$$(X(3915)\rightarrow J/\psi\omega)$ is ($52\pm10\pm3)$ eV and $(10.5\pm1.9\pm0.6)$ eV for the spin hypotheses $J=0$ and $J=2$, respectively, where the first error is statistical and the second is systematic. In addition, a Bayesian upper limit (UL) at 90\% confidence level (CL) is obtained for the X(3872), $\Gamma_{\gamma\gamma}[X(3872)]$ x $\cal{B}$$(X(3872)\rightarrow J/\psi\omega)<$ 1.7 eV, assuming J=2.

\section {Search for the $Z_1(4050)^+$ and $Z_2(4250)^+$}
\begin{figure}[htb]
  \centering
  \includegraphics[width=80mm]{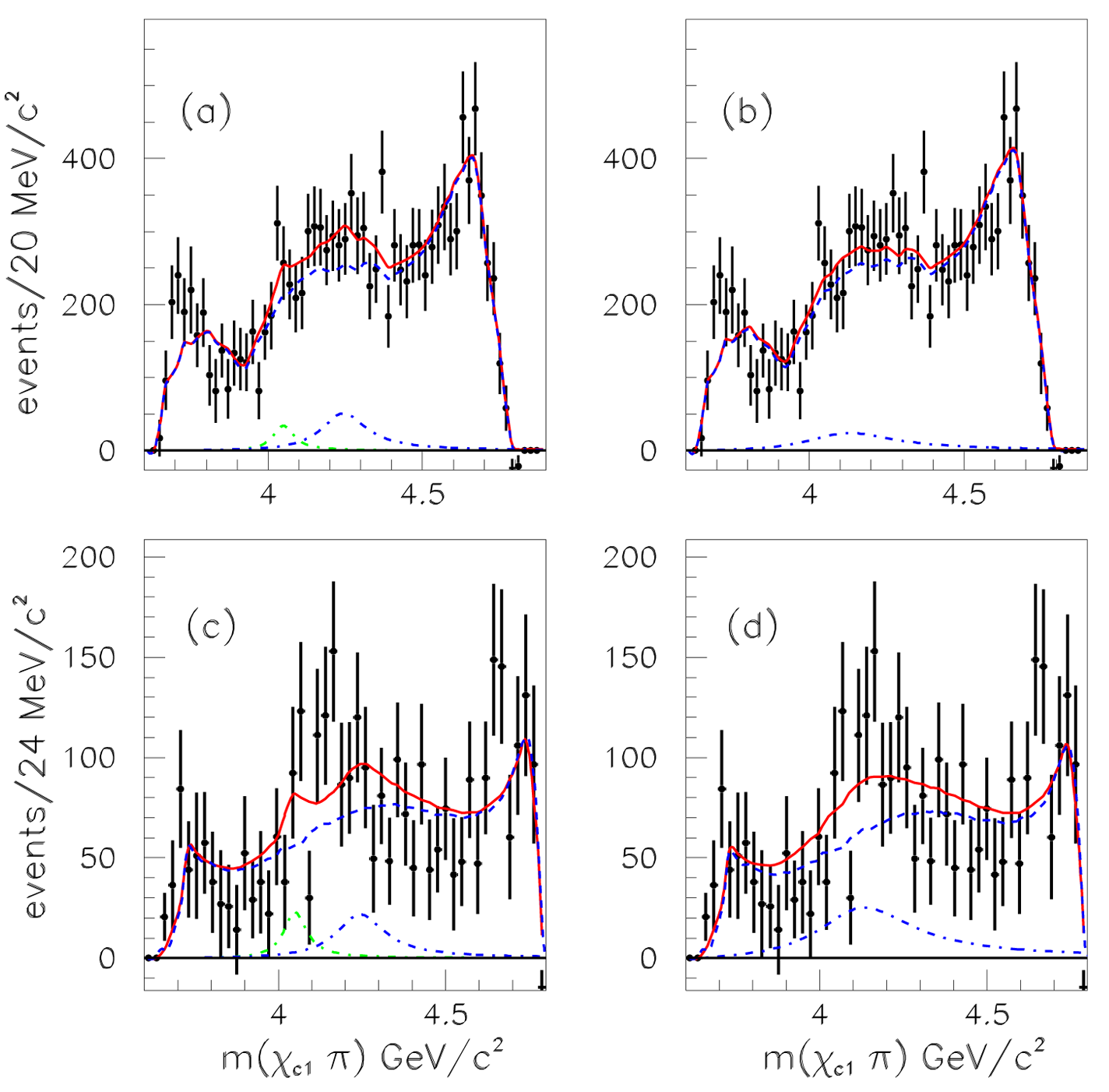}
  \caption{\scriptsize Fit to the background-subtracted and efficiency-corrected $\chi_{c1}\pi$ mass distributions. See text for more details.}
  \label{Fig:antimo}
\end{figure}

Belle reported the observation of two resonance-like structures, $Z_1(4050)^+$ and $Z_2(4250)^+$ in the study of $\bar{B}^0\rightarrow\chi_{c1}K^-\pi^+$, both decaying to $\chi_{c1}\pi^+$ \cite{belleZ}. \\
\indent \babar studied the same final states \cite{antimo} to search for the $Z_1(4050)^+$ and $Z_2(4250)^+$ decay into $\chi_{c1}\pi^+$ in $\bar{B}^0\rightarrow\chi_{c1}K^-\pi^+$ and $B^+\rightarrow K_S^0\chi_{c1}\pi^+$ (this mode is analyzed for the first time) where $\chi_{c1}\rightarrow J/\psi\gamma$, using a data sample of 429 fb$^{-1}$. The $\chi_{c1}\pi^+$ mass distribution, background-subtracted and efficiency-corrected, was modeled using the $K\pi$ mass distribution and the corresponding normalized $K\pi$ Legendre polynomial moments. Figure \ref{Fig:antimo} shows the results of the fits to the $\chi_{c1}\pi^+$ mass spectra. The fit shown in Figure \ref{Fig:antimo}(a) includes both $Z_1(4050)^+$ and $Z_2(4250)^+$ resonances and the fit shown in Figure \ref{Fig:antimo}(b) includes a single broad $Z(4150)^+$ resonance. Figures \ref{Fig:antimo}(c,d) show the $\chi_{c1}\pi$ mass spectrum fitted in the Dalitz plot region 1.0 $\le$ $m^2(K\pi)<$ 1.75 GeV$^2$/c$^4$ in order to make a direct comparison to the Belle results \cite{belleZ} (this region is labeled as "window" in Table \ref{Tab:antimo}). The results of the fits are summarized in Table \ref{Tab:antimo} and in every case the yield significance does not exceed 2 $\sigma$. The ULs on the 90\% CL on the branching fractions are: $\cal{B}$$(\bar{B}^0\rightarrow Z_1(4050)^+K^-)$ x $\cal{B}$$(Z_1(4050)^+\rightarrow\chi_{c1}\pi^+)<$ 1.8 x 10$^{-5}$; $\cal{B}$$(\bar{B}^0\rightarrow Z_2(4250)^+K^-)$ x $\cal{B}$$(Z_2(4250)^+\rightarrow\chi_{c1}\pi^+)<$ 4.0 x 10$^{-5}$ and $\cal{B}$$(\bar{B}^0\rightarrow Z^+K^-)$ x $\cal{B}$$(Z^+\rightarrow\chi_{c1}\pi^+)<$ 4.7 x 10$^{-5}$.

\begin{table}[ht]
\caption{\scriptsize Results of the $\chi_{c1}\pi$ fits. $N_\sigma$ and Fraction give, for each fit, the significance and the fractional contribution of the Z resonances.}
\label{Tab:antimo}
\centering
{\small
\begin{tabular}{lllll}
\hline
Data & Resonances & $N_{\sigma}$ & Fraction (\%) & $\chi^2/NDF$ \\
\hline
a) Total & $Z_1(4050)^+$ & 1.1 & 1.6$\pm$1.4 & 57/57\\
              & $Z_2(4250)^+$ & 2.0 & 4.8$\pm$2.4 & \\
b) Total & $Z(4150)^+$ & 1.1 & 4.0$\pm$3.8 & 61/58\\
\hline
a) Window & $Z_1(4050)^+$ & 1.2 & 3.5$\pm$3.0 & 53/46\\
              & $Z_2(4250)^+$ & 1.3 & 6.7$\pm$5.1 & \\
b) Window & $Z(4150)^+$ & 1.7 & 1.37$\pm$8.0 & 53/47\\
\hline
\end{tabular}
}
\end{table}

\section {Study of the $J/\psi\pi^+\pi^-$ system via Initial State Radiation (ISR)}
\begin{figure}[htb]
  \centering
  \includegraphics[scale=0.8]{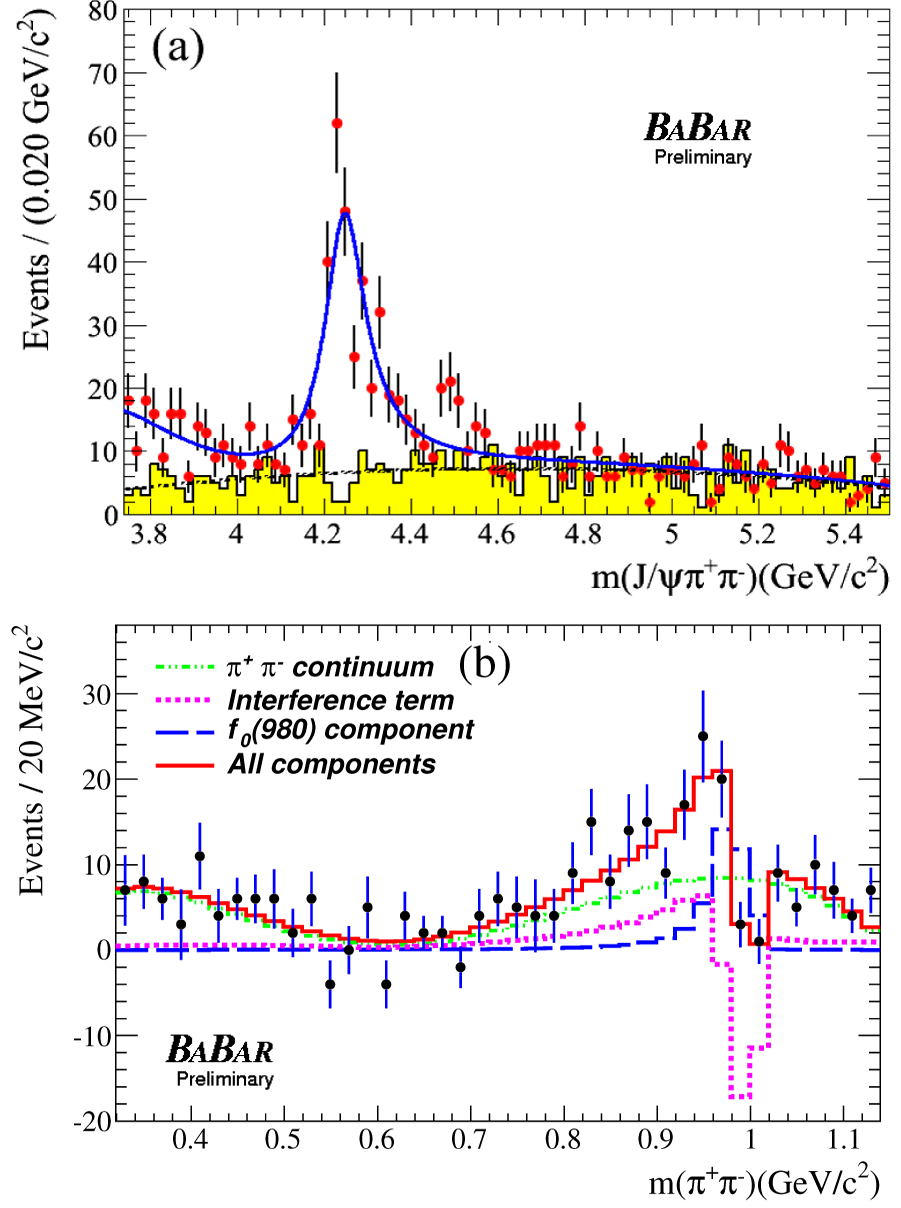}
  \caption{\scriptsize (a) The fit to the $J/\psi\pi^+\pi^-$ invariant mass distribution. (b) The fit to the dipion invariant mass distribution from the Y(4260) signal region.}
  \label{Fig:vale1}
\end{figure}

The Y(4260) charmonium-like resonance was discovered by \babar \cite{Y} in ISR production of $J/\psi\pi^+\pi^-$. A subsequent Belle analysis \cite{Ybelle} of the same final state suggested also the existence of an additional resonance around 4.1 GeV/c$^2$ that they dubbed the Y(4008).\\
\indent  The \babar analysis of the $J/\psi\pi^+\pi^-$ system produced in ISR has been repeated using a data sample of 454 fb$^{-1}$ \cite{valentina}. \\
\indent  The $J/\psi\pi^+\pi^-$ mass region below $\sim$4 GeV/c$^2$ is investigated for the first time. 
In that region an excess of events has been observed and the conclusion, after a detailed study of the $\psi(2S)$ lineshape (to estimate the $\psi(2S)$ tail contribution to that region), is that it is not possible to discount the possibility of a contribution from a $J/\psi\pi^+\pi^-$ continuum cross section in this region. From this study we obtain the cross section value 14.05 $\pm$ 0.26 (stat) pb for radiative return to the $\psi(2S)$ and a measurement of the width $\Gamma(\psi(2S)\rightarrow e^+e^-)=2.31\pm0.05$ (stat) keV. Figure \ref{Fig:vale1}(a) shows the fit to the $J/\psi\pi^+\pi^-$ distribution. A clear signal of the Y(4260) is observed for which the values obtained are $m[Y(4260)]=4244\pm5\pm4$ MeV/c$^2$, $\Gamma[Y(4260)]=114^{+16}_{-15}\pm7$ MeV and $\Gamma_{ee}$ x $\cal{B}$$(J/\psi\pi^+\pi^-)=9.2\pm0.8$ (stat) $\pm$ 0.7 (syst) eV. No evidence for the state at $\sim$ 4 GeV/c$^2$ reported by Belle \cite{Ybelle} was seen. A study of the $\pi^+\pi^-$ system from the Y(4260) decay to $J/\psi\pi^+\pi^-$ has been performed. The dipion system is in a predominantly S-wave state. The mass distribution exhibits an $f_0(980)$ signal, for  which a simple model indicates a branching ratio with respect to $J/\psi\pi^+\pi^-$ of 0.17 $\pm$ 0.13 (stat). The fit to the dipion invariant mass distribution is shown in Figure \ref{Fig:vale1}(b).

\section {Study of the $\psi(2S)\pi^+\pi^-$ system via ISR}

\begin{figure}[htb]
  \centering
  \includegraphics[scale=0.23]{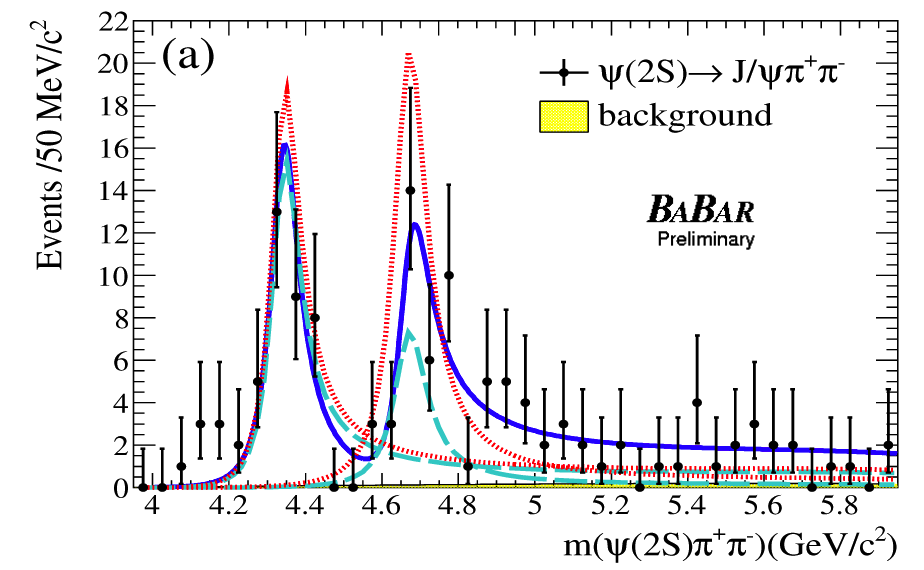}
  \caption{\scriptsize Invariant mass distribution of the $\psi(2S)\pi^+\pi^-$ system}
  \label{Fig:vale2}
\end{figure}

The Y(4350) charmonium-like resonance was discovered by \babar \cite{Y4350} in ISR production of $\psi(2S)\pi^+\pi^-$. Belle, in \cite{Y4660} reported also the existence of an additional resonance around 4.660 GeV/c$^2$.\\
\indent The \babar analysis studies the $\psi(2S)\pi^+\pi^-$ system produced via ISR using the full datasets collected at the $\Upsilon(nS)$, n=2,3,4; this corresponds to an integrated luminosity of 520 fb$^{-1}$. \\
\indent An unbinned extended-maximum-likelihood fit is performed to the $\psi(2S)\pi^+\pi^-$ invariant mass distribution and simultaneously to the background distribution. The result of the fit is shown in Figure \ref{Fig:vale2}.

In Table \ref{Tab:reson} the mass and width values obtained for the two resonances are reported.

\begin{table}[ht]
\caption{\scriptsize Mass and width values for the states observed in the $\psi(2S)\pi^+\pi^-$ ISR process (\babar preliminary).}
\label{Tab:reson}
\centering
{\small
\begin{tabular}{lll}
\hline
Resonance & Mass (MeV/c$^2$) & $\Gamma$ (MeV) \\
\hline
Y(4360) & 4340$\pm$16$\pm$9 & 94$\pm$32$\pm$13\\
Y(4660) & 4669$\pm$21$\pm$3 & 104$\pm$48$\pm$10\\
\hline
\end{tabular}
}
\end{table}

\end{document}